\pdfoutput=1
\documentclass[11pt,twoside,a4paper,cmspaper,final,collab]{cms-tdr}
\def\svnVersion{61a15ff}\def\svnDate{2025/11/19}\def\cmsCernNoTag{CERN-EP-2024-310}\def\cmsCernDate{\today}\def\cmsMessage{Published in Physical Review Letters as \href{http://dx.doi.org/10.1103/51fw-klz3}{\doi{10.1103/51fw-klz3}.}}
\begin{document}\cmsNoteHeader{SUS-23-015}

\newcommand{\mH}{\ensuremath{m_{\PSGcpmDo}}\xspace}
\newcommand{\mHtwo}{\ensuremath{m_{\PSGczDt}}\xspace}
\newcommand{\mL}{\ensuremath{m_{\PSGczDo}}\xspace}
\newcommand\ttZ{\ensuremath{\ttbar\PZ}\xspace}
\newcommand{\ttW}{\ensuremath{\ttbar\PW}\xspace}
\newcommand\WZ{\ensuremath{\PW\PZ}\xspace}
\newcommand{\ZZ}{\ensuremath{\PZ\PZ}\xspace}
\newcommand{\ZG}{\ensuremath{\PZ\PGg}\xspace}
\newcommand{\JHUGen}{{\textsc{JHUGen}}\xspace}
\newcommand{\ST}{\ensuremath{S_{\mathrm{T}}}\xspace}
\newcommand{\Nj}{\ensuremath{N_{\mathrm{j}}}\xspace}
\newcommand{\Nb}{\ensuremath{N_{\mathrm{\PQb}}}\xspace}
\newcommand{\dR}{\ensuremath{\Delta R}\xspace}
\newcommand{\dRmin}{\ensuremath{\Delta R_{\text{min}}}\xspace}
\newcommand{\nossf}{\ensuremath{{\text{OSSF}}n}\xspace}
\newcommand{\mmin}{\ensuremath{m_{\text{min}}}\xspace}
\newcommand{\R}{\ensuremath{R}}
\newcommand{\SUtwo}{\ensuremath{\text{SU}(2)_{\text{L}}}\xspace}
\newcommand{\Uone}{\ensuremath{\text{U}(1)_{\text{Y}}}\xspace}

\cmsNoteHeader{SUS-23-015}
\title{Search for new physics in jet multiplicity patterns of multilepton events at \texorpdfstring{$\sqrt{s} = 13\TeV$}{sqrt(s) = 13 TeV}}

\author*[cern]{CMS Collaboration}
\date{\today}

\abstract{
A first search for beyond the standard model physics in jet multiplicity patterns of multilepton events is presented, using a data sample corresponding to an integrated luminosity of 138\fbinv of 13\TeV proton-proton collisions recorded by the CMS detector at the LHC. The search uses observed jet multiplicity distributions in one-, two-, and four-lepton events to explore possible enhancements in jet production rate in three-lepton events with and without bottom quarks. The data are found to be consistent with the standard model expectation. The results are interpreted in terms of supersymmetric production of electroweak chargino-neutralino superpartners with cascade decays terminating in prompt hadronic \R-parity violating interactions.
}

\hypersetup{
pdfauthor={CMS Collaboration},
pdftitle={Search for new physics in jet multiplicity patterns of multilepton events at sqrt(s) = 13 TeV},
pdfsubject={CMS},
pdfkeywords={CMS, multilepton, multijet, SUSY, RPV, electroweak production, LSP, hadronic decay}
}

\maketitle

{\tolerance=800
Following the discovery of the Higgs boson~\cite{CMS:2012qbp,ATLAS:2012yve}, a number of critical questions remain open outside of the standard model~(SM) of particle physics, such as physics underlying electroweak~(EW) symmetry breaking and the Higgs boson mass~\cite{Coleman:1973jx,Farzinnia:2013pga,Rossi:2023ara}, the origin of neutrino masses~\cite{Minkowski:1977sc,Magg:1980ut,Foot:1988aq,Brdar:2019iem}, the identity of dark matter, and the observed baryon asymmetry in the Universe~\cite{Kaplan:1991ah,Jungman:1995df,Buchmuller:2005eh,Hooper:2011ti}.
These make the search for new physics beyond the SM~(BSM) an important mission of the CERN LHC.
Despite the wide array of theoretical models put forth for BSM physics, the guidance they offer is not as specific as it was for the discovery of the latest pieces of the SM, like the top quark and the Higgs boson, and no direct signs of new physics have been found yet.
Therefore, experimental efforts at the LHC must adopt a broad approach in searching for signs of BSM physics.
\par}

A compelling subset of BSM theories features events with multiple final-state leptons and enhanced production of hadronic jets in high energy proton-proton collisions~\cite{Choudhury:2024ggy,Alvarez:2020ffi,vonBuddenbrock:2018xar,Beuria:2017jez,Bhattacherjee:2009jh}.
The primary SM backgrounds for such signatures include production of massive vector bosons and top quarks in association with high jet multiplicity arising from initial- (ISR) and final-state radiation~(FSR).
The characterization of these SM background contributions faces major challenges as predictions of differential jet multiplicity distributions generally suffer from large theoretical uncertainties, including insufficient modeling of higher-order quantum chromodynamics~(QCD) and EW effects.
Significant discrepancies exist in state-of-the-art event and particle shower generators, as exemplified in recent studies of \PW, \PZ, \ttbar, \WZ, {\ZZ}, and \ttZ processes at the LHC~\cite{ATLAS:2022nrp,ATLAS:2019hxz,ATLAS:2019bsc,ATLAS:2023eld,ATLAS:2017bcd,CMS:2010svw,CMS:2022ilp,CMS:2021vhb,CMS:2024ybg,CMS:2021icx,CMS:2019too,CMS:2024ilq}.

These shortcomings impose serious limitations on experimental searches for BSM physics distinguished by such telltale hadronic enhancements. 
Indeed, analyses by the ATLAS and CMS Collaborations frequently rely on inclusive selection criteria and observables, instead of exploiting the exclusive multiplicity of jets~\cite{ATLAS:2021wob,ATLAS:2022yhd,ATLAS:2023sbu,ATLAS:2023lfr,ATLAS:2023afl,ATLAS:2020uer,CMS:2020cpy,CMS:2021cox,CMS:2019hsm,CMS:2019lwf,CMS:2022nty,CMS:2024xdq}.
Variables such as the transverse momentum~(\pt) imbalance due to invisible particles, transverse masses of invisible and visible particles, and scalar sum of transverse momenta of visible particles are less affected by these theoretical uncertainties, but they provide diminished sensitivity to jet production due to new physics.
Furthermore, three-lepton events consistent with on-shell \PZ and \PW boson decays are often used as data control regions~(CR) for determining corrections to production cross sections for simulations of \WZ and \ttZ backgrounds.
Thus, signatures that resemble such SM processes but with an enhancement in jet multiplicity are not sufficiently covered in BSM search efforts.

This Letter presents the first direct search for BSM physics in exclusive jet multiplicity distributions of proton-proton collision events containing three leptons.
The analysis targets events with on-shell, leptonically decaying \PW and \PZ bosons accompanied by several soft jets.
Accurate predictions of jet distributions from SM \WZ and \ttZ processes are thus essential for detecting such new physics.
Previous studies~\cite{Ellis:1985vn,Berends:1989cf,Giele:1989id,Berger:2010vm,Gerwick:2011tm,Gerwick:2012hq,Bern:2014fea,Tricoli:2020uxr} have shown that jet production in events with a hard process, such as energetic lepton production, follows patterns like ``staircase'' and ``Poisson'' scaling.
Despite dependence on jet and event selection, jet multiplicity distributions are expected to show approximate universality with common patterns within similar process families.
The three-lepton \WZ and \ttZ processes belong to two separate such families within the SM background; \WZ is primarily quark initiated with ISR-driven jet production, while \ttZ is dominantly gluon initiated with jets from ISR, FSR, and top-quark decays.

In this Letter, we employ a novel method that exploits expected similarities in jet production across SM processes within the same family to probe BSM scenarios that manifest as deviations in exclusive jet multiplicity distributions, particularly through additional jets from massive particle decays.
Specifically, we derive the \WZ and \ttZ jet distributions in three-lepton events, where BSM-enhanced jet activity is posited, using relatively signal-depleted two- and one-lepton data events enriched in SM processes of the same family.
This data-driven approach substantially improves the modeling of associated hadronic activity, mitigating the related uncertainties via the use of SM-like CRs and significantly enhancing sensitivity to BSM-induced jet activity.

The results are used to probe a previously unexplored class of supersymmetric~(SUSY)~\cite{Wess:1973kz,Wess:1974tw,Nilles:1983ge,Barbieri:1982eh,Haber:1984rc,Dawson:1983fw} models that predict events with \PW and \PZ bosons along with an enhanced jet multiplicity. 
EW production of a chargino-neutralino pair $\PSGcpmDo \PSGczDt$ (with masses $\mH=\mHtwo$) yields a \PW and \PZ boson through cascade decays to a pair of the lightest SUSY particle~(LSP) neutralino \PSGczDo (with mass \mL), $\PSGcpmDo  \to \PW^{\pm} \PSGczDo$ and $\PSGczDt \to \PZ \PSGczDo$.
Each LSP decays promptly afterward via hadronic R-parity violating~(RPV) interactions~\cite{Hall:1983id,Ross:1984yg,Dreiner:1997uz,Barbier:2004ez,Barman:2020azo}.
The search targets three-lepton events from on-shell leptonic decays of the massive vector bosons, $\PW \to \Pell \nu$ and $\PZ \to \Pell \Pell$, and we consider two scenarios for the three-quark decay of the LSP, $\PSGczDo \to \PQu\PQd\PQs$ or $\PSGczDo \to \PQu\PQd\PQb$, referred to as RPVq and RPVb, respectively.
The chargino pair $\PSGcpDo\PSGcmDo$ production does not contribute to the three-lepton signal, and its contribution to the one- and two-lepton CRs is negligible for this class of SUSY models.

The analysis is based on a data sample of 138\fbinv of 13\TeV $\Pp\Pp$ collisions recorded by the CMS detector in 2016--2018~\cite{CMS:2021xjt,CMS:2018elu,CMS:2019jhq}.
The CMS apparatus~\cite{CMS:2008xjf} is a nearly hermetic multipurpose detector, designed to trigger on~\cite{CMS:2020cmk,CMS:2016ngn} and identify electrons, muons, photons, and hadrons~\cite{CMS:2020uim,CMS:2018rym,CMS:2019ied,CMS:2014pgm}.
A global ``particle-flow'' algorithm~\cite{CMS:2017yfk} is used to reconstruct all individual particles in an event.
It combines information provided by the all-silicon inner tracker and by the crystal electromagnetic and brass-scintillator hadron calorimeters, operating inside a 3.8\unit{T} superconducting solenoid, with data from the gas-ionization muon detectors embedded in the flux-return yoke outside the solenoid.
The reconstructed particles are used to build jets and missing transverse momentum (\ptmiss)~\cite{CMS:2016lmd,CMS:2019ctu}.

Jets are clustered using the anti-\kt algorithm with a radius parameter of 0.4, as implemented in the \textsc{FastJet} package~\cite{Cacciari:2008gp,Cacciari:2011ma,CMS-PAS-JME-16-003}, and their energy is corrected on a jet-by-jet basis~\cite{CMS-DP-2020-019}.
Jets originating from {\PQb} quarks are identified with the \textsc{DeepJet} algorithm operated at a medium working point having 70\% average efficiency, with 10\% and 1\% efficiencies for misidentifying charm and light quarks, respectively~\cite{Sirunyan:2017ezt,Bols:2020bkb,CMS-DP-2018-058}.
The vector \ptvecmiss is defined as the projection onto the plane perpendicular to the beam axis of the negative vector-momenta sum of all reconstructed particle-flow objects in an event.

The data used in this search were recorded using a combination of isolated single-electron and -muon triggers with corresponding electron and muon \pt thresholds of 27 and 24\GeV in 2016, 32 and 27\GeV in 2017, and 32 and 24\GeV in 2018, respectively.
We select events that contain one to four leptons (electrons or muons) and any number of jets.
The leptons are required to be well isolated with $\pt>10\GeV$ and pseudorapidity $\abs{\eta}<2.4$~\cite{CMS:2018rym,CMS:2020uim}.
At least one of the leptons must satisfy tighter criteria with $\pt>30\,(35)\GeV$ in 2016 (2017--2018) for an electron or with $\pt>26\,(29)\GeV$ in 2016 and 2018 (2017) for a muon.
All jets are required to satisfy $\pt>30\GeV$ and $\abs{\eta}<2.5$, as well as distance in $\eta$-$\phi$ space~\cite{CMS:2008xjf} $\dR\equiv[(\Delta\eta)^2+(\Delta\phi)^2]^{1/2}>0.4$ from a selected muon or electron, where $\phi$ is the azimuthal angle in radians.
In order to suppress contributions from low-mass quarkonium resonances, cascade decays of heavy-flavor hadrons, and low-\dR FSR, events with two or more leptons are discarded if $\mmin<12\,\GeV$ or $\dRmin<0.2$, where \mmin and \dRmin are defined, respectively, as the minimum invariant mass and minimum \dR of all lepton pairs in the event, regardless of lepton charge and flavor.

Events satisfying the baseline selections are categorized using a set of invariant mass and multiplicity variables.
The number of distinct opposite-sign same-flavor~(OSSF) lepton pairs in an event is denoted as \nossf, where $n$ is the number of such pairs, such that a $\PGmp\PGmm\PGmm$ event is labeled OSSF1 and a $\PGmp\PGmp\PGmm\PGmm$ event is labeled OSSF2.
Events with an OSSF lepton pair with invariant mass between 76 and 106\GeV, consistent with leptonic decays of the \PZ boson, are labeled OnZ.
Otherwise, they are labeled OffZ.
The minimum lepton \pt requirement is raised to 20\GeV in all three-lepton OnZ events.

\begin{figure*}[ht]
\centering
\includegraphics[width=0.315\textwidth]{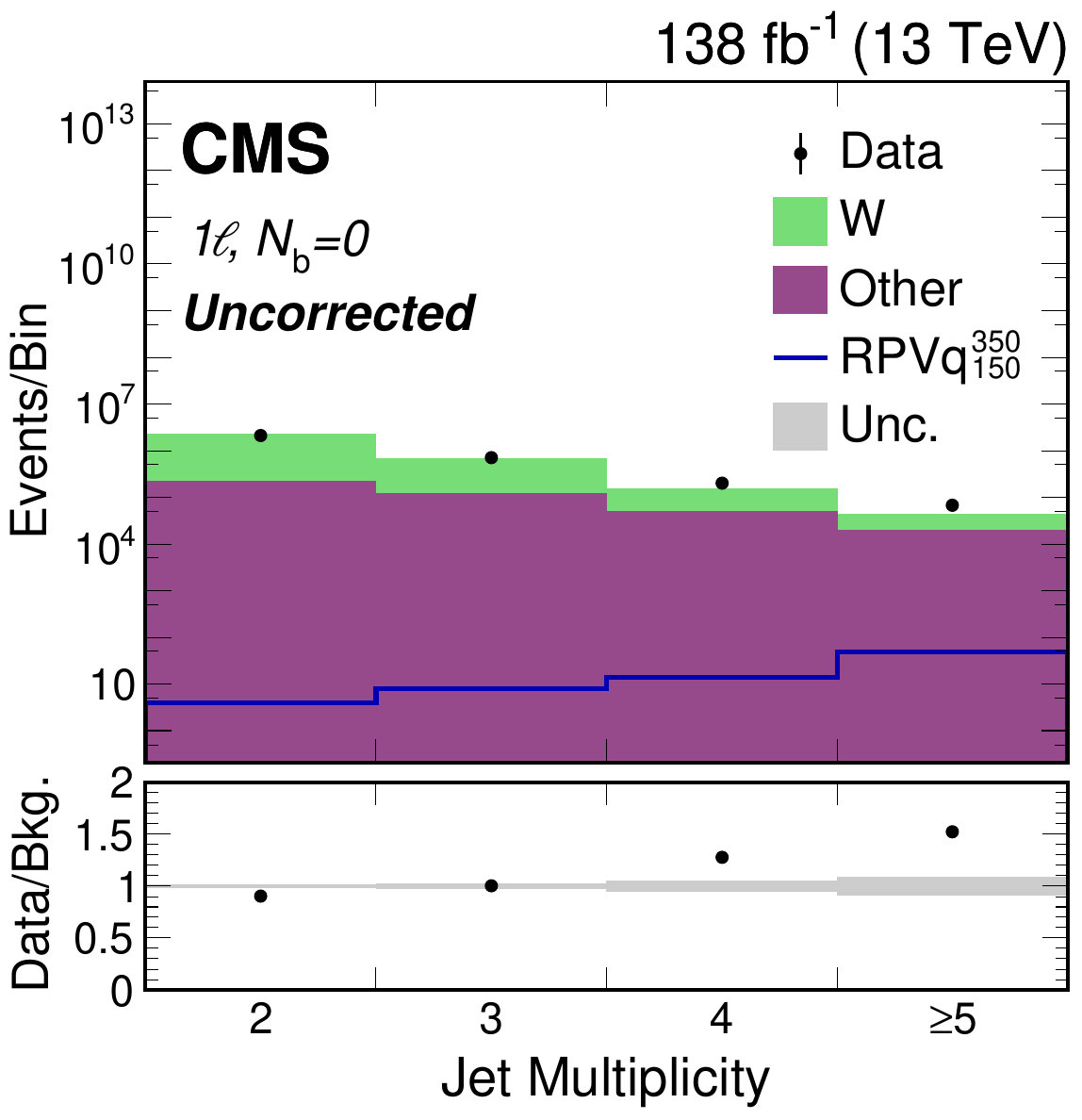}
\includegraphics[width=0.315\textwidth]{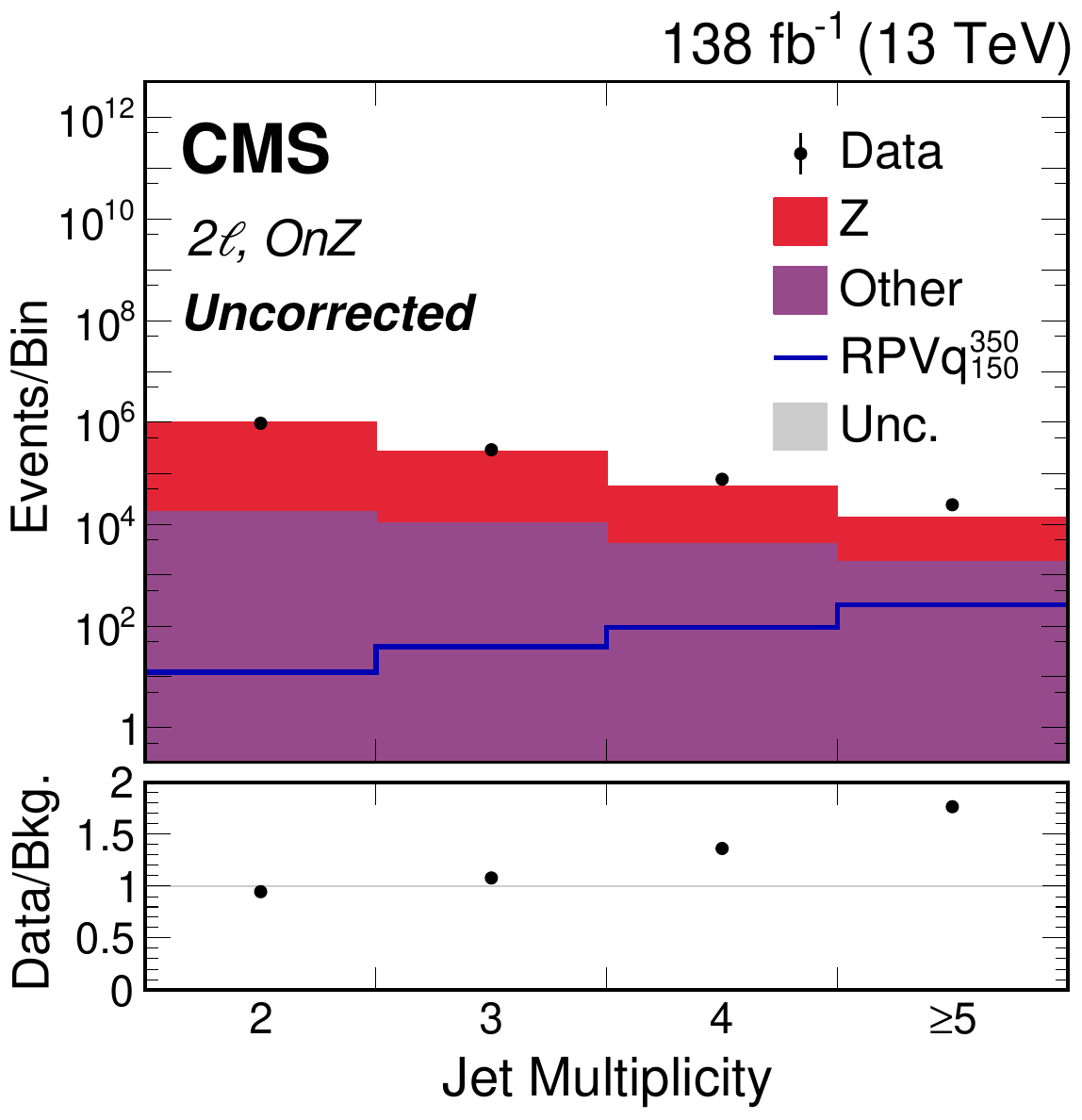}
\includegraphics[width=0.315\textwidth]{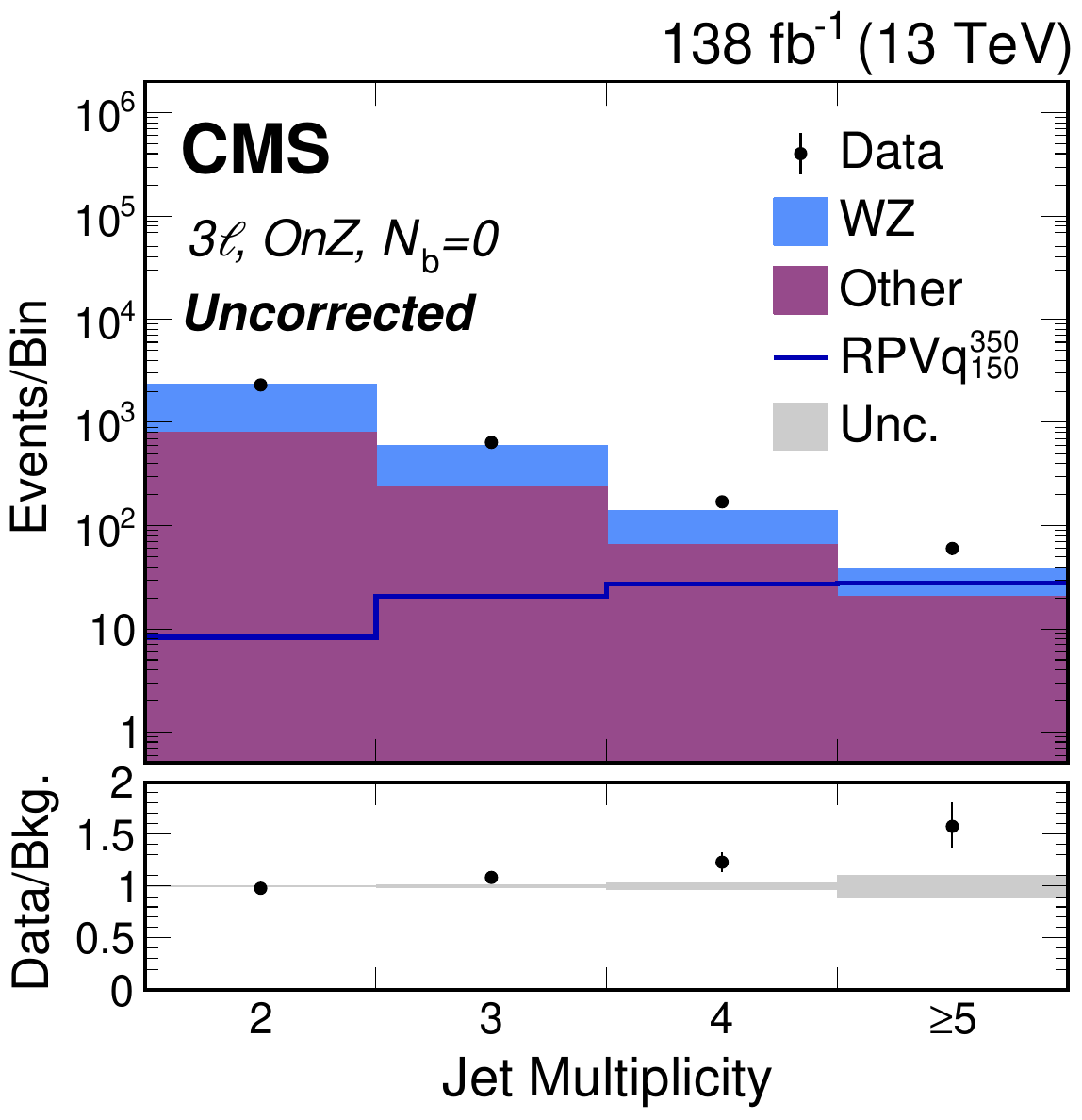}
\caption{\label{fig:Njet}
Distributions of jet multiplicity in a \PW-enriched selection of one-lepton events (left), a \PZ-enriched selection of two-lepton OnZ events (center), and a \WZ-enriched selection of three-lepton OnZ events (right, search region) prior to corrections.
The lower panels show the ratio of observed events (Data) to the total background prediction (Bkg.), with the gray bands representing the statistical uncertainty (Unc.) in the predictions.
SM background predictions fall short of the observations in all cases for higher jet multiplicities.
Expected RPVq ($\mH=\mHtwo=350\GeV$, $\mL=150\GeV$) signal distributions are overlaid for comparison.
}
\end{figure*}

Figure~\ref{fig:Njet} shows the jet multiplicity distributions across one-, two-, and three-lepton events enriched in \PW, \PZ, and \WZ contributions, respectively.
These indicate the impact, size, and similarity of the jet multiplicity mismodeling across different processes as well as the shortcoming of the simulated \WZ sample in higher jet-multiplicity bins.
In particular, these discrepancies in three-lepton events should be taken into account when interpreting potential signal-like features, such as those of the RPVq signal.
This constitutes the primary motivation for the data-driven jet multiplicity correction and the choice of the signal model.

The search regions~(SRs) comprise the three-lepton OnZ events with at least two jets. 
These are primarily classified based on their jet~(\Nj) and \PQb{}-tagged jet~(\Nb) multiplicities.
We define eight nonoverlapping SRs, namely with $\Nj=2$, $\Nj=3$, $\Nj=4$, and $\Nj\geq5$, further split according to $\Nb=0$ or $\Nb\geq1$.
The remaining one-, two-, three-, and four-lepton events are used as signal-depleted CRs to isolate, characterize, and quantify the major irreducible SM background processes, and to tailor the simulations to improve the description of SM processes.
Although RPVq and RPVb signals can produce up to six jets, scenarios with low LSP mass or with Lorentz-boosted LSPs produce fewer jets in LSP decays because of the requirements on the reconstructed jet \pt and jet radius parameters, populating SRs with lower \Nj and \Nb values.
In each \Nj and \Nb SR, a number of \ST bins are defined to probe for wide enhancements above SM expectations, where \ST is an event-based effective mass variable defined as \ptmiss plus the scalar \pt sum of electrons, muons, and jets.
The \ST scales with the total mass of the BSM particles and mitigates the model-specific combinatorial problem of building multibody resonances.
The splitting of SRs along dimensions of jet multiplicity and \ST provides additional sensitivity against SM processes which may produce similar numbers of jets in the final state but with a different overall effective mass. 

In all leptonic events, electrons and muons produced from the decays of \PW, \PZ, and Higgs bosons (either directly or via a leptonically decaying \Pgt lepton) are referred to as prompt leptons, and are often indistinguishable in momentum and isolation from those produced in signal events.
Thus, the SM contributions consisting of only prompt leptons are referred to as the irreducible backgrounds.
Reducible backgrounds are defined as those which include jets misidentified as leptons or where the leptons originate from heavy-quark decays, and background processes with such leptons are collectively labeled as ``MisID'' backgrounds.

Simulated samples are used in the estimation of the irreducible background as well as signal contributions.
All SM samples were generated using a matrix-element calculation at fixed order in QCD, with up to a maximum of two additional hard-scattering jets, depending on the process.
The \WZ, \ttZ, \ttW, \ZG, and triboson backgrounds were generated using \MGvATNLO 2.6.5~\cite{Alwall:2014hca} at next-to-leading order~(NLO) precision.
The background contribution from \ZZ production via quark-antiquark annihilation was generated using \POWHEG 2.0~\cite{Nason:2004rx,Frixione:2007vw,Alioli:2010xd} at NLO, while the contribution from gluon-gluon fusion production was generated at leading order~(LO) using \MCFM 7.0.1~\cite{Campbell:2010ff}.
Processes from Higgs boson production were generated using \POWHEG, \MGvATNLO, and \JHUGen 7.0.11~\cite{Gao:2010qx,Bolognesi:2012mm,Anderson:2013afp,Gritsan:2016hjl} at NLO.
Additionally, \PZ, \PW, and \ttbar processes primarily used for various studies in one- and two-lepton events were also generated at NLO with \MGvATNLO.
Contributions from other processes involving single or multiple top quarks or vector bosons were simulated using \MGvATNLO at LO or NLO in QCD.
All signal samples were generated with \MGvATNLO at LO precision.
A simplified model spectrum~\cite{Alwall:2008ag,LHCNewPhysicsWorkingGroup:2011mji} is assumed for the class of SUSY models targeted in this work, consisting of degenerate winolike chargino and neutralino that are superpartners of the \SUtwo gauge fields with masses $\mH=\mHtwo\geq 125\GeV$, together with a binolike neutralino LSP that is the superpartner of the \Uone gauge field with mass $\mL\leq \mH-100\GeV$~\cite{Fayet:1976cr,Martin:1997ns}.
All other SUSY particles are assumed to be sufficiently decoupled to be inaccessible at the LHC.

The parton showering, fragmentation, and hadronization of the matrix element level events in all background and signal samples were performed with \PYTHIA 8.240~\cite{Sjostrand:2014zea} with the event tune CP5~\cite{CMS:2019csb}.
The FxFx~\cite{Frederix:2012ps} or MLM~\cite{Alwall:2007fs} matching scheme is used in \MADGRAPH samples to remove double-counted partons between the matrix element calculations and parton shower description, at NLO or LO precision, respectively.
The NNPDF3.1 next-to-NLO parton distribution functions~(PDFs) were used to generate all samples~\cite{Ball:2017nwa}.
The response of the CMS detector was simulated using the \GEANTfour toolkit~\cite{Agostinelli:2002hh}.
Aside from the hard-scattering event, additional $\Pp\Pp$ interactions from within the same or adjacent bunch crossing~(pileup) are simulated and incorporated in the samples, and simulated events are reweighted according to the pileup distribution in the recorded data.

\begin{figure*}[ht]
\centering
\includegraphics[width=0.315\textwidth]{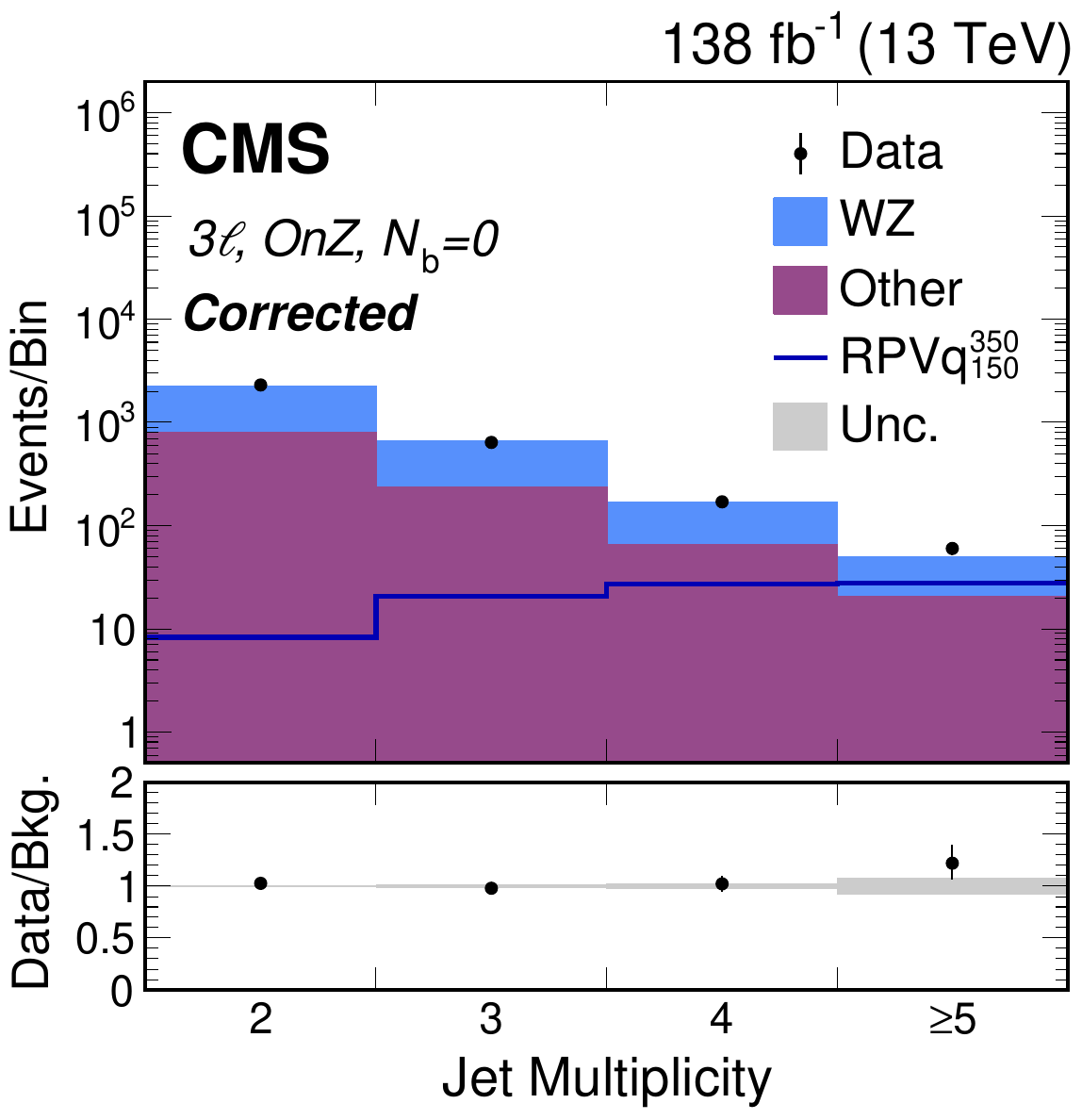}
\includegraphics[width=0.315\textwidth]{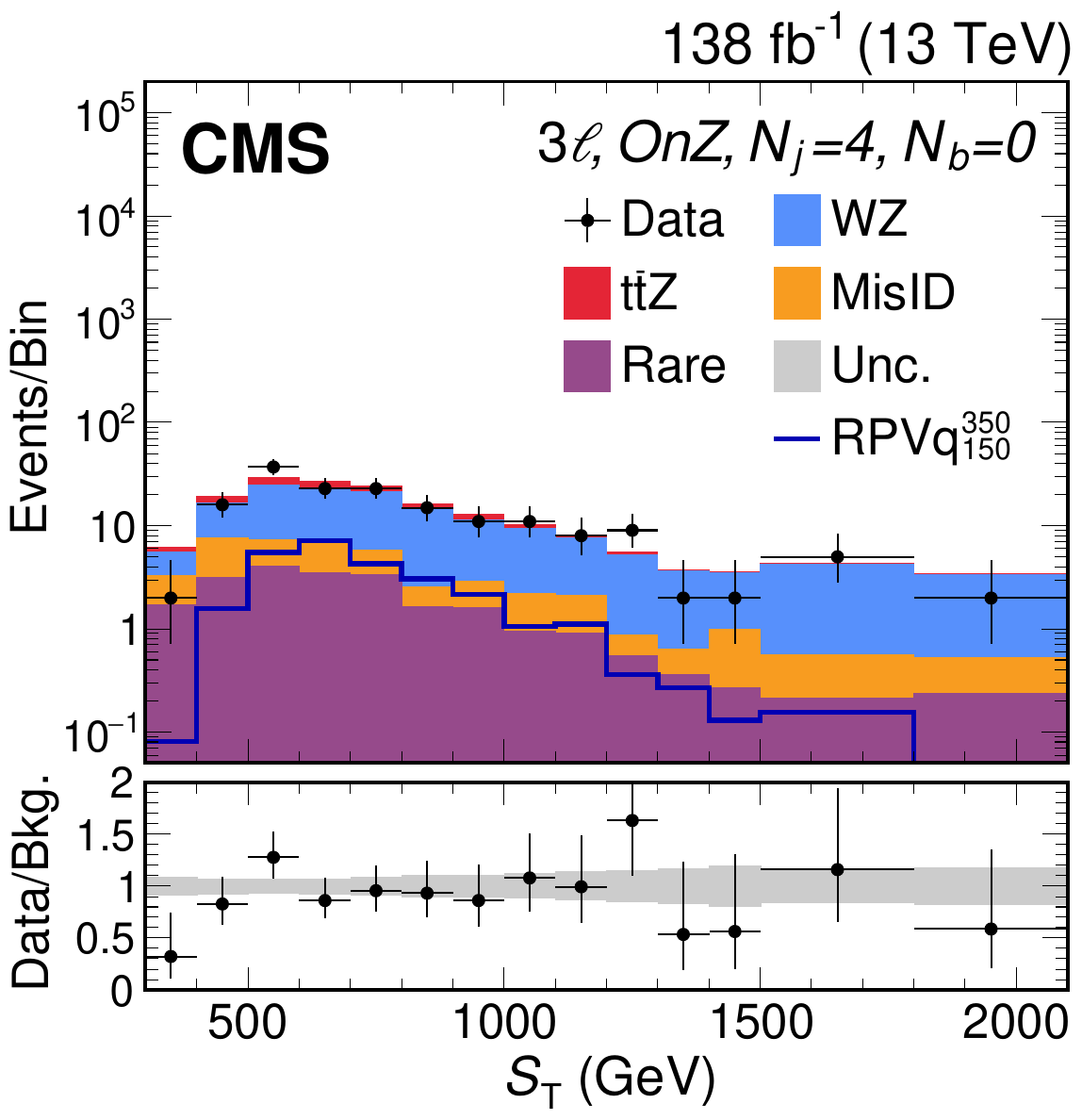}
\includegraphics[width=0.315\textwidth]{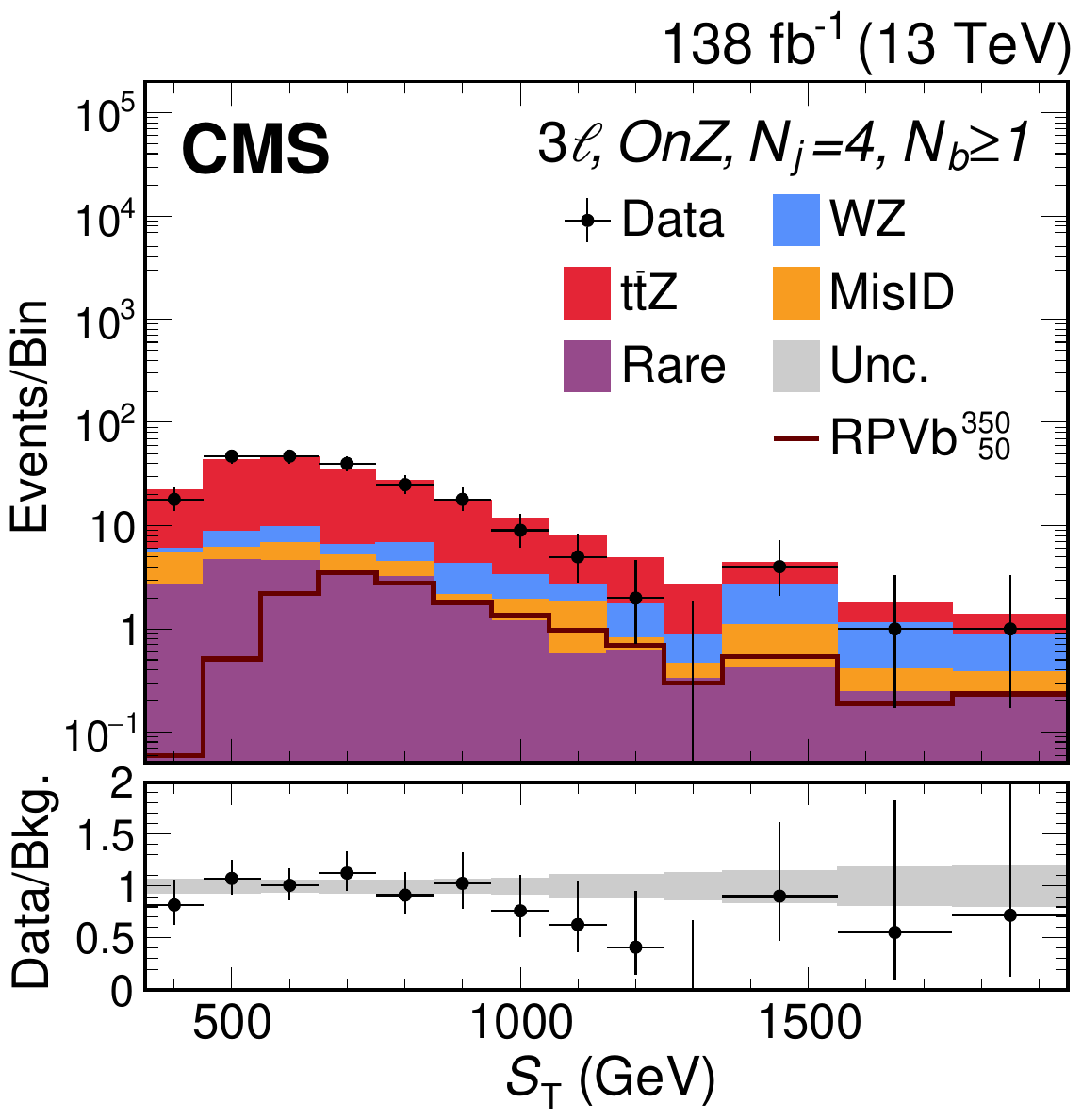}
\caption{\label{fig:SR}
Example search region distributions of jet multiplicity with $\Nb=0$ (left) as well as \ST in four-jet bins with $\Nb=0$ (center) and $\Nb\geq1$ (right). 
The lower panels show the ratio of observed events (Data) to the total background prediction (Bkg.), with the gray bands representing the sum of statistical (left), or statistical and systematic (center and right) uncertainties (Unc.) in the predictions.
The expected background jet multiplicity distribution is obtained after applying the data driven corrections, whereas the \ST distributions and the uncertainties are obtained after also performing the binned maximum likelihood fit to the data under the background-only hypothesis.
Expected RPVq and RPVb ($\mH=\mHtwo=350\GeV$ with $\mL=150$ and 50\GeV, respectively) signal distributions are overlaid for comparison.
}
\end{figure*}

The \WZ and \ttZ processes are the dominant irreducible SM backgrounds for $\Nb=0$ and $\Nb\geq1$ SR events, respectively.
A major limitation in simulating their differential jet multiplicity distributions arises from mismodeling of QCD showering and the choice of factorization and renormalization scales.
To address these effects in SRs ($\Nj\geq2$), we exploit similarities in jet production mechanisms between the three-lepton \WZ and \ttZ processes and the two-lepton \PZ and one-lepton \ttbar processes, respectively.
The latter yield significantly larger event samples in data and are used to define CRs with minimal signal contamination.
Data-driven differential jet multiplicity distributions from these CRs are applied as shape corrections to the \WZ and \ttZ simulations, systematically defining and constraining the ratios of exclusive jet multiplicity bins.

A two-lepton selection of \PZ-enriched events is used to derive corrections for the three-lepton \WZ background, in which the accompanying jets arise predominantly from ISR.
Although both \PZ and \WZ production proceed via EW interactions, differences in the invariant mass of the hard-scattering system affect the production characteristics of the recoiling jets.
These differences are found to be largely ameliorated by an additional selection of $\pt > 100\GeV$ on the dilepton system (reconstructed \PZ boson candidate) in the \PZ-enriched events.
The method is validated both in simulated samples, as well as in a one-lepton data CR enriched in \PW boson contributions, where applying the \PZ-derived corrections reduces discrepancies between data and simulation to below 10\% for $\Nj\geq2$.
Figure~\ref{fig:SR} (left) shows the corrected \Nj distribution in the \WZ-dominated three-lepton SR, where the data are observed to broadly align with the expected SM background.

Similarly, a one-lepton selection of \ttbar-enriched events is used to derive corrections for the three-lepton \ttZ background.
Both processes involve ISR, FSR, and jets from top decays ($\PQt \to \PW\PQb \to \PQq\PQq\PQb$), and have comparable hard-scattering scales.
The compatibility of their jet multiplicity behavior is validated in simulation, with no significant differences observed.
Moreover, this compatibility is verified in data using two- and four-lepton CRs enriched in \ttbar and \ttZ events, respectively.

The shape of the \ST distribution in each SR jet-multiplicity bin is obtained from simulation.
Selections of \PZ- and \ttbar-enriched two-lepton CR events are used to validate \ST modeling performance in exclusive jet multiplicity bins, which is observed to be consistent with data up to the individual normalization factors.
A set of three- and four-lepton CRs are further defined to verify the modeling of \ST, dilepton mass, and other relevant kinematic variables, as well as to obtain cross section normalization factors for the irreducible backgrounds.
For the \WZ contributions, a selection of three-lepton OnZ events with 0 or 1 jet is used, whereas a \ttZ-enriched CR is created by selecting four-lepton events with only one OnZ pair, at least one \PQb-tagged jet, and  $\ST>350\GeV$.
Other contributions mostly come from \ZG and \ZZ processes, where the former contributes via ISR or FSR photons that convert asymmetrically such that only one of the resultant leptons is reconstructed in the detector, and the latter contributes via events where one of the leptons originating from the \PZ boson decays is not reconstructed or identified as an electron or muon.
A dedicated OffZ three-lepton selection is defined for the \ZG process, where the total mass of the three-lepton system is required to be within the \PZ boson mass window of $91\pm15\GeV$.
Four-lepton OSSF2 events with either two OnZ pairs and any number of \PQb-tagged jets, or one OnZ pair and no \PQb-tagged jets, are used to normalize and study \ZZ simulations.
Additionally, dedicated jet-multiplicity-dependent corrections are derived for the \ZZ background simulation using the signal-depleted four-lepton events. 
Other SM processes without a dedicated CR, including triboson, Higgs boson, and other rare SM contributions, are normalized to their NLO or LO theoretical cross sections.
Contributions from these processes together with relatively small contributions from  \ZZ and \ZG processes are collectively referred to as ``Rare'' backgrounds.
The final cross section normalizations of all major irreducible backgrounds and signal are ultimately simultaneously obtained by a binned maximum likelihood fit to all SRs and three- and four-lepton CRs, where the normalized cross sections from the CRs are solely used as initial values.

The MisID background contributions are estimated using a ``matrix method''~\cite{ATLAS:2022swp,CMS:2015nep} from a set of sideband regions defined in data for each SR. 
The sideband regions consist of events with leptons passing less stringent selection criteria, where at least one lepton fails the isolation requirements used in the analysis.
The method used in this analysis follows the implementation detailed in Refs.~\cite{CMS:2022nty,CMS:2024ulc}, and is validated in OnZ three-lepton simulated events as well as signal-depleted OSSF0 and OSSF1 OffZ three-lepton data events.

A number of systematic uncertainties arise from the corrections applied to the background and signal simulations.
These include corrections for the efficiencies of the electron and muon triggers, lepton reconstruction, identification, and isolation, as well as the lepton energy scale and resolution modeling, and for jets and \ptmiss, \PQb tagging efficiency, pileup modeling, and energy scale and resolution corrections.
Each of these is considered for the main irreducible \WZ and \ttZ backgrounds and the signal samples, and evaluated by varying the corresponding correction factor up and down by one standard deviation of its associated uncertainty.
The resulting shape variations in the \ST spectra of the main irreducible backgrounds in exclusive \Nj and \Nb bins are used to define associated systematic uncertainties, which are typically less than 10\%.
The uncertainties in the integrated luminosity vary in the range of 1.2--2.5\% among the data-taking years~\cite{CMS:2021xjt,CMS:2018elu,CMS:2019jhq}.
Uncertainties due to the NNPDF3.1 PDF set~\cite{Ball:2017nwa}, the choices of factorization and renormalization scales~\cite{Cacciari:2003fi}, and the ISR and FSR parton shower parameters~\cite{CMS:2019csb} used in the \WZ and \ttZ samples are also evaluated, and while subdominant, are still considered.

The MisID background estimate from control samples in data has an uncertainty in the range of 20--30\%~\cite{CMS:2022nty,CMS:2024ulc}.
The \ZZ background has a dominant set of uncertainties, in the range of 15--50\%, for the exclusive jet-multiplicity-dependent corrections derived from the dedicated four-lepton CR data events.
For the rare background processes, a 50\% systematic uncertainty is assigned to the theoretical cross section estimates to cover any higher-order effects and PDF uncertainties.
These have a minor impact on the results as the corresponding processes are not major contributors in the SRs.
Additionally, all background and signal estimates have uncertainties from the limited numbers of events in simulated samples or data sidebands.
These statistical uncertainties are typically less important, but are nonetheless propagated to the results.

The dominant systematic uncertainties arise from the normalization of the \WZ and \ttZ backgrounds in the $N_j\geq 2$ phase space as well as the uncertainties applied to the \WZ and \ttZ jet multiplicity corrections.
The normalization factors for these backgrounds and the targeted signal process are included as free parameters in the maximum likelihood fit.
The jet multiplicity corrections affect the \Nj shapes of the \WZ and \ttZ contributions, and are implemented as independent constraints on the ratios of event counts in neighboring jet multiplicity bins.
A relative uncertainty of 30\% is assigned to each of these ratio constraints as induced by various systematic effects, where the primary contributors are the residual nonclosure effects in the modeling of \WZ and \ttZ processes by \PZ and \ttbar processes as studied in simulated samples (up to 20\%) and the stability of correction factors as functions of data-taking period and detector conditions (up to 10\%).
In the one-lepton \ttbar and two-lepton \PZ-enriched data events where the jet multiplicity corrections are measured, the contributions of other processes are estimated and subtracted using simulated samples.
Thus, we also evaluate the impact of various simulation-based uncertainties on the corrections.
Among these, jet energy scale and \PQb tagging efficiency correction uncertainties as well as normalization uncertainties are the most relevant, but their total impact on the jet multiplicity corrections is measured to be less than 5\%. 
In contrast, the statistical uncertainties  are found to be negligible throughout.

\begin{figure*}[htb!]
\centering
\includegraphics[width=0.45\textwidth]{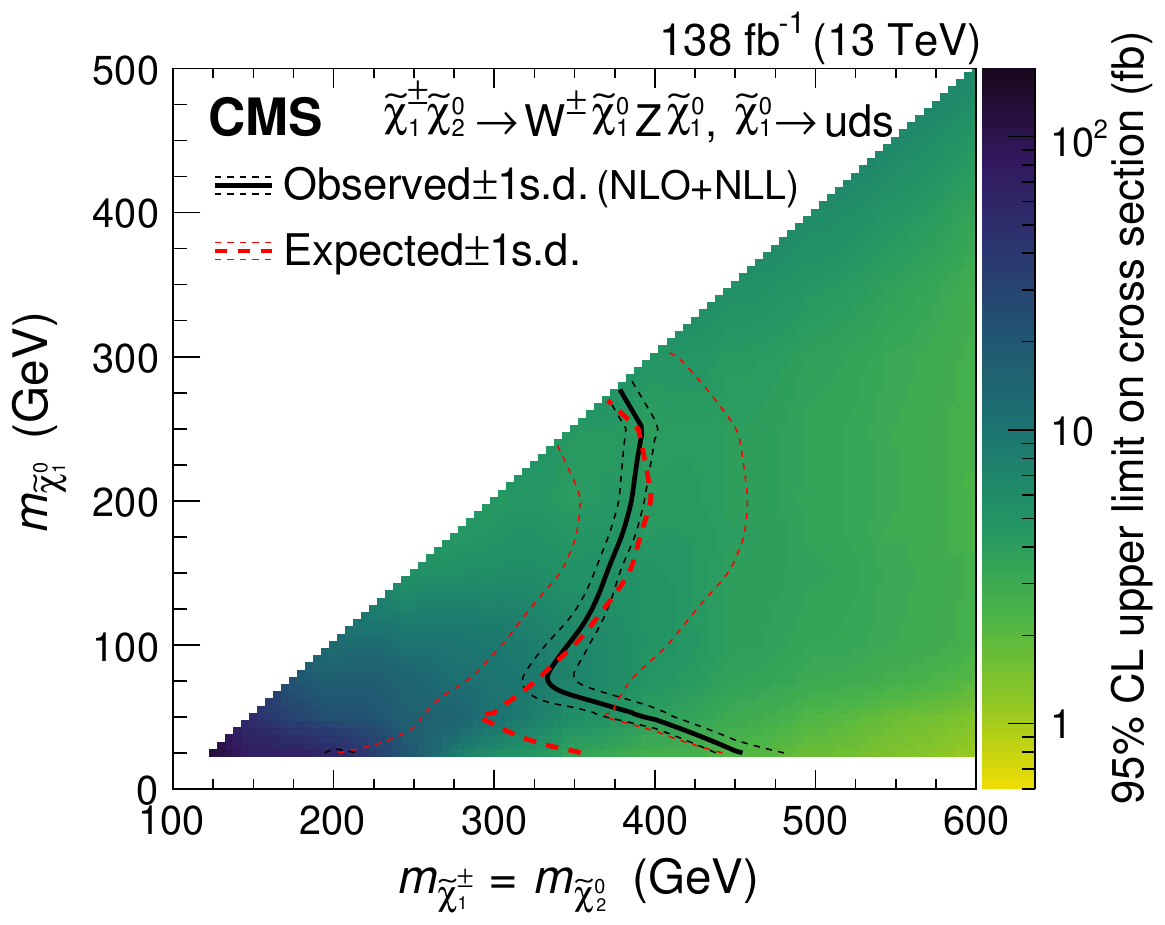}
\includegraphics[width=0.45\textwidth]{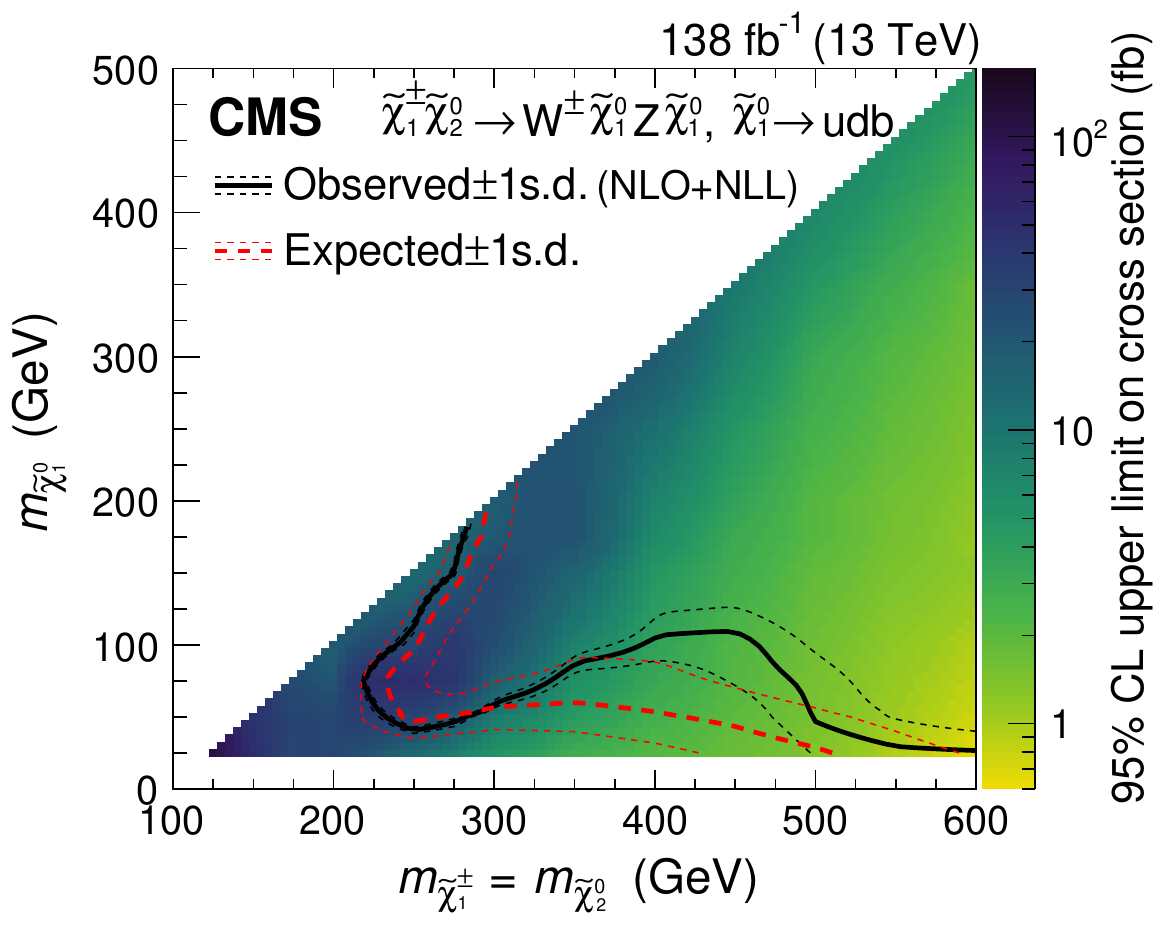}
\caption{\label{fig:limits}
The 95\% confidence level upper limits on the RPVq (left) and RPVb (right) signal production cross sections as a function of $\mH=\mHtwo$ and \mL.
Contour lines indicate the observed (bold solid) and median expected (bold dashed) boundaries with 1 standard deviation (s.d.) as well as the 68\% expected bands (thin dashed), where the latter on the observed contour line is due to the theoretical uncertainty on the signal production cross section.
}
\end{figure*}

The following results are determined using the CMS statistical analysis tool, \textsc{Combine}~\cite{CMS:2024onh}. 
A simultaneous binned maximum likelihood fit is performed over 13 bins in three- and four-lepton CRs and 106 jet-multiplicity and \ST bins in SRs.
Sources of systematic uncertainties, including those of statistical origin, are incorporated in the likelihood as nuisance parameters.
These include unconstrained parameters for the normalization factors of \WZ, \ttZ, \ZZ, \ZG, and MisID background contributions, as well as log-normal constrained parameters for all other systematic uncertainties.
The statistical uncertainties in the signal and background estimates are modeled with gamma functions.
No statistically significant deviation from the SM background is observed.
Representative \ST background spectra in four-jet SRs are shown in Fig.~\ref{fig:SR} with $\Nb=0$ (center), and $\Nb\geq 1$ (right).

Using the modified frequentist approach with the \CLs criterion~\cite{Junk:1999kv,Read:2002hq} in the asymptotic approximation~\cite{Cowan:2010js}, the results are interpreted as upper limits at 95\% confidence level on the production cross section of chargino-neutralino pairs in the class of SUSY models explored here, separately considering the RPVq and RPVb signal scenarios.
The signal production cross sections are computed at NLO plus next-to-leading logarithmic~(NLL) precision under the simplified model assumption~\cite{Fuks:2012qx,Fuks:2013vua}. 
Figure~\ref{fig:limits} shows the observed and expected upper limits on the excluded $\PSGcpmDo \PSGczDt$ production cross sections.
In the range of probed chargino-neutralino masses, the RPVq (RPVb) model is excluded for an LSP mass of up to 275 (180)\GeV.
Tabulated results are provided in the associated HEPData record~\cite{hepdata}.

Previously, the ATLAS and CMS Collaborations have set constraints on \R-parity conserving or RPV scenarios with chargino-neutralino pair production in signatures involving low hadronic activity or significant \pt imbalance due to undetected stable neutralino LSPs~\cite{ATLAS:2024qxh,CMS:2024gyw}, or in signatures with three or more charged leptons~\cite{ATLAS:2020uer,CMS:2016zgb}, as well as constraints on long-lived RPV decays of metastable neutralino LSPs~\cite{ATLAS:2011zmp,ATLAS:2023oti,CMS:2016vuw,CMS:2021tkn} and hadronic RPV decays of superpartners involving top quarks~\cite{ATLAS:2021fbt,CMS:2021knz}.
In comparison, these are the first dedicated constraints targeting EW superpartner pair production with prompt decays of the LSPs via hadronic RPV interactions in final states with jets and three leptons.

In summary, the first search for beyond the standard model physics in jet multiplicity distributions of multilepton events has been performed using proton-proton collision data collected with the CMS detector at $\sqrt{s} = 13\TeV$, corresponding to an integrated luminosity of 138\fbinv.
No significant deviations from standard model expectations have been observed and the first direct bounds are set on a class of supersymmetric models populating final states with jets and three leptons.

\begin{acknowledgments}
We congratulate our colleagues in the CERN accelerator departments for the excellent performance of the LHC and thank the technical and administrative staffs at CERN and at other CMS institutes for their contributions to the success of the CMS effort. In addition, we gratefully acknowledge the computing centers and personnel of the Worldwide LHC Computing Grid and other centers for delivering so effectively the computing infrastructure essential to our analyses. Finally, we acknowledge the enduring support for the construction and operation of the LHC, the CMS detector, and the supporting computing infrastructure provided by the following funding agencies: SC (Armenia), BMBWF and FWF (Austria); FNRS and FWO (Belgium); CNPq, CAPES, FAPERJ, FAPERGS, and FAPESP (Brazil); MES and BNSF (Bulgaria); CERN; CAS, MoST, and NSFC (China); MINCIENCIAS (Colombia); MSES and CSF (Croatia); RIF (Cyprus); SENESCYT (Ecuador); ERC PRG, RVTT3 and MoER TK202 (Estonia); Academy of Finland, MEC, and HIP (Finland); CEA and CNRS/IN2P3 (France); SRNSF (Georgia); BMBF, DFG, and HGF (Germany); GSRI (Greece); NKFIH (Hungary); DAE and DST (India); IPM (Iran); SFI (Ireland); INFN (Italy); MSIP and NRF (Republic of Korea); MES (Latvia); LMTLT (Lithuania); MOE and UM (Malaysia); BUAP, CINVESTAV, CONACYT, LNS, SEP, and UASLP-FAI (Mexico); MOS (Montenegro); MBIE (New Zealand); PAEC (Pakistan); MES and NSC (Poland); FCT (Portugal); MESTD (Serbia); MICIU/AEI and PCTI (Spain); MOSTR (Sri Lanka); Swiss Funding Agencies (Switzerland); MST (Taipei); MHESI and NSTDA (Thailand); TUBITAK and TENMAK (Turkey); NASU (Ukraine); STFC (United Kingdom); DOE and NSF (USA).
\end{acknowledgments}

\bibliography{auto_generated} 
\cleardoublepage \appendix\section{The CMS Collaboration \label{app:collab}}\begin{sloppypar}\hyphenpenalty=5000\widowpenalty=500\clubpenalty=5000
\cmsinstitute{Yerevan Physics Institute, Yerevan, Armenia}
{\tolerance=6000
A.~Hayrapetyan, A.~Tumasyan\cmsAuthorMark{1}\cmsorcid{0009-0000-0684-6742}
\par}
\cmsinstitute{Institut f\"{u}r Hochenergiephysik, Vienna, Austria}
{\tolerance=6000
W.~Adam\cmsorcid{0000-0001-9099-4341}, J.W.~Andrejkovic, T.~Bergauer\cmsorcid{0000-0002-5786-0293}, S.~Chatterjee\cmsorcid{0000-0003-2660-0349}, K.~Damanakis\cmsorcid{0000-0001-5389-2872}, M.~Dragicevic\cmsorcid{0000-0003-1967-6783}, P.S.~Hussain\cmsorcid{0000-0002-4825-5278}, M.~Jeitler\cmsAuthorMark{2}\cmsorcid{0000-0002-5141-9560}, N.~Krammer\cmsorcid{0000-0002-0548-0985}, A.~Li\cmsorcid{0000-0002-4547-116X}, D.~Liko\cmsorcid{0000-0002-3380-473X}, I.~Mikulec\cmsorcid{0000-0003-0385-2746}, J.~Schieck\cmsAuthorMark{2}\cmsorcid{0000-0002-1058-8093}, R.~Sch\"{o}fbeck\cmsAuthorMark{2}\cmsorcid{0000-0002-2332-8784}, D.~Schwarz\cmsorcid{0000-0002-3821-7331}, M.~Sonawane\cmsorcid{0000-0003-0510-7010}, W.~Waltenberger\cmsorcid{0000-0002-6215-7228}, C.-E.~Wulz\cmsAuthorMark{2}\cmsorcid{0000-0001-9226-5812}
% ...
\par}
\vskip\cmsinstskip
\dag:~Deceased\\
$^{1}$Also at Yerevan State University, Yerevan, Armenia\\
$^{2}$Also at TU Wien, Vienna, Austria\\
% ...
\end{sloppypar}
\end{document}